\documentclass[aps,prl,twocolumn,superscriptaddress,showpacs,showkeys,amsmath,amssymb]{revtex4-1}

\usepackage{amsfonts}
\usepackage{amssymb,amsmath}
\usepackage{graphicx}
\usepackage{bm}
\usepackage{color}
\newcommand{\abs}[1]{\left|#1\right|}

\begin{document}

\title{Flat-Band Ferromagnetism as a Pauli-Correlated Percolation Problem}

\author{M. Maksymenko}
\affiliation{Institute for Condensed Matter Physics, National Academy of Sciences of Ukraine,
        1 Svientsitskii Street, 79011 L'viv, Ukraine}
\affiliation{Max-Planck-Institut f\"{u}r Physik Komplexer Systeme,
        N\"{o}thnitzer Stra{\ss}e 38, 01187 Dresden, Germany}
\author{A. Honecker}
\affiliation{Institut f\"ur Theoretische Physik,
 Georg-August-Universit\"at G\"ottingen, 37077 G\"ottingen, Germany}
\affiliation{Fakult\"at f\"ur Mathematik und Informatik,
             Georg-August-Universit\"at G\"ottingen, 37073 G\"ottingen, Germany}
\author{R. Moessner}
\affiliation{Max-Planck-Institut f\"{u}r Physik Komplexer Systeme,
        N\"{o}thnitzer Stra{\ss}e 38, 01187 Dresden, Germany}
\author{J. Richter}
\affiliation{Institut f\"{u}r Theoretische Physik, Universit\"{a}t Magdeburg,
        P.O. Box 4120, 39016 Magdeburg, Germany}
\author{O. Derzhko}
\affiliation{Institute for Condensed Matter Physics, National Academy of Sciences of Ukraine,
        1 Svientsitskii Street, 79011 L'viv, Ukraine}
\date{September 14, 2012}



\begin{abstract}
We investigate the location and nature of the para-ferro transition of 
interacting electrons in dispersionless bands using the example of the 
Hubbard model on the Tasaki lattice. This case can be analyzed as a 
geometric site-percolation problem where different configurations appear 
with nontrivial weights. We provide a complete exact solution for the 1D 
case and develop a numerical algorithm for the 2D case. In two dimensions the 
paramagnetic phase persists beyond the uncorrelated percolation point, and 
the grand-canonical transition is via a first-order jump to an {\em 
unsaturated} ferromagnetic phase.
\end{abstract}

\maketitle

{\it{Introduction.}}---The interplay of the Coulomb interaction with the Pauli principle was
already recognized by Heisenberg \cite{WHeisenberg} 
to give rise to a ferromagnetic exchange interaction, also encoded in Hund's
rule about aligned spins in a partially filled shell. For a many-body
system of correlated electrons with a flat band, when the interaction
energy completely dominates over the kinetic energy, the ferromagnetic
instability is one of the few problems for which exact results are
available, albeit for a restricted range of fillings 
\cite{mielke, tasakiPRL, tasakiCMP, sawtooth1,sawtooth2, 1dtasaki_cerh3b2}.

Flat band systems are receiving a great deal of attention right now, 
in particular with the view of realizing new many-body phases there
(see \cite{optical1,alter_1,topological, topological_tasaki,organic_polymers} and references 
therein); in this context, the possibility of ferromagnetism as a 
many-body instability is also being considered 
\cite{neupert_mudry}. It is therefore timely to provide a 
detailed study of the phase diagram and the critical properties of this form of magnetism: 
we analyze a flat-band ferromagnet with an on-site Hubbard
interaction of strength $U \ge 0$. For $U=0$, any
state involving electrons occupying the flat band only is trivially a ground
state. 

Crucially, this degeneracy is only partially lifted when a repulsive
$U>0$ is switched on.
First, since the flat band permits well-localized real-space electronic wave 
functions, at low density electrons can be placed on the lattice so that they do not overlap. 
Second, even if they do overlap, they can still avoid paying an energy penalty $U$: the 
basic reason is that the Pauli principle, by demanding an antisymmetric
pair wave function, makes the overlap between two electrons on the same site
vanish provided  they are in a symmetric spin state. This is the origin of flat-band ferromagnetism. 

As the density of
electrons increases, ferromagnetic clusters of increasing size appear. The
degeneracy, $m+1$, of a ferromagnetic cluster containing $m$ electrons,
gives differing weights to different clustering of electrons. The
ferromagnetic transition corresponds to the emergence of a cluster
containing a nonzero fraction of the electrons.

An early remark by Mielke \cite{mielke} likened this problem to one of percolation. Mielke
and Tasaki \cite{tasakiPRL,tasakiCMP} noted that, for a class of flat-band ferromagnets on
particular decorated lattices, the percolation problem in question is not
a standard one \cite{stauffer,isichenko} but rather one including nontrivial weights.

Here, we develop this analogy in detail. First of all, we point
out that the interaction between the clusters, on account of its
``statistical origin'' in the Pauli principle, is unusual in that it is
range-free and purely geometric---two particles interact only if they form
part of the same cluster. The interaction is genuinely many-body 
in that it cannot be decomposed into a 
sum of pairwise terms. It is effectively repulsive and only
depends on the size of the cluster, irrespective of its 
shape. Despite its long range, the 
statistical interaction does saturate.

This motivates the study of the resulting unusual percolation problem, which we call
Pauli-correlated percolation (PCP). We find that it has a number of interesting
features in its own right. It provides an instance of a problem in the
quantum physics of strongly correlated electrons which can be ``reduced'' to a
highly nontrivial problem in {\em classical} statistical mechanics, on which an
entirely different set of tools can be brought to bear.
\begin{figure*}
\centering
\includegraphics[clip=on,width=\textwidth, angle=0]{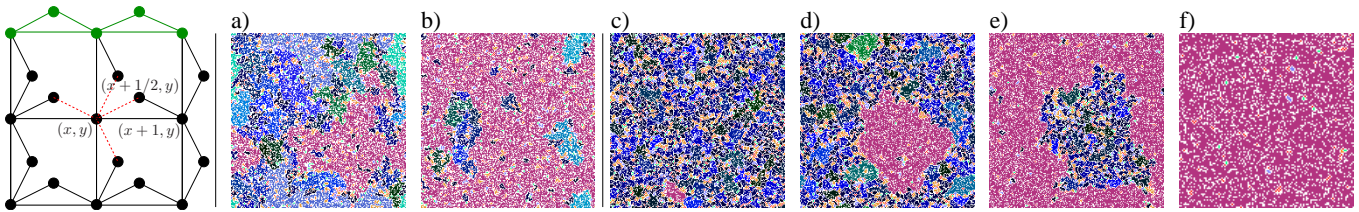}
\caption
{(Color)  Left: Two-dimensional Tasaki lattice. A trapping cell contains five sites (dashed red lines).
The green circles  and lines show the 1D variant of the lattice (sawtooth chain). Right: Snapshots of configurations for 
standard and Pauli-correlated percolation for small deviations from critical concentration.
Panels a) and b) show snapshots (lattice extension ${\cal{L}}=200$) of configurations for standard percolation 
for concentrations $p_{1}=0.574$ and $p_{2}=0.6$ ($p_{c}=0.592746\ldots$), while
panels c), d), e), and f) show snapshots 
for Pauli-correlated percolation for $p_{3}=0.62$ (paramagnetic),
$p_{4}=0.65$, $p_{5}=0.7$ (phase-separated), and $p_{6}=0.78$
(ferromagnetic). Pink color denotes the 
largest cluster.
}
\label{fig01}
\end{figure*}
We first demonstrate some special features of this problem by providing a
complete exact solution of the one-dimensional (1D) version of this model, which corresponds to a sawtooth lattice 
potentially realized in strongly correlated sawtoothlike 
compounds such as $\rm{CeRh_{3}B_{2}}$ \cite{1dtasaki_cerh3b2}. Unlike standard percolation, this  
exhibits a tendency to break up large clusters as well as a development
of spatial (anti)correlations. Its percolation transition at full filling is continuous.

Next, we carry out an analysis of the  phase diagram for the two-dimensional (2D) Tasaki lattice, a 
decorated square lattice (see left panel of Fig. 1). Using a numerical 
algorithm custom-tailored to the problem at hand by extending the Hoshen-Kopelman and Newman-Ziff algorithms \cite{hoshen,ziff}
for standard percolation, we establish that the
ferromagnetic transition does indeed take place at a filling comfortably
in excess of the corresponding  well-known percolation transition on
the square lattice at $p_c=0.592746\ldots$ \cite{isichenko,ziff}.
In the grand-canonical ensemble, this transition is of first order;
in the canonical ensemble we find concomitant phase-separated states
(see Fig.\ \ref{fig01} for some examples). 

{\it{Pauli-correlated percolation and flat-band ferromagnetism.}}---As 
a representative system with a dispersionless (flat) band,
let us consider the
%
Tasaki model \cite{tasakiPRL,tasakiCMP}, although our approach 
in principle can be adapted to other flat-band lattices. The
enumeration of all ground states
of the repulsive Hubbard model on the Tasaki lattice maps to a
percolation problem where each occupied site on a hypercubic
lattice corresponds to an electron localized in a trapping cell 
(whose wave function only overlaps with that of 
electrons in adjacent cells.) (The details of this mapping, which are 
unimportant for the following, are relegated to section \ref{sec:SupplPcp}).
All ground states can be labeled by the possible geometric configurations
of $n$ electrons distributed over $\cal{N}$ traps, labeled by $q$,
and a nontrivial weight of each state \cite{tasakiCMP}
\begin{equation}
\label{degeneracy}
W(q)=\prod_{i=1}^{M_q} \, e^{\mu\,\abs{C_i}}\, (\abs{C_i}+1)~~,
\end{equation}
which arises because of the spin degeneracy of the
ferromagnetic cluster of size $\abs{C_i}$
in configuration $q$ ($M_q$ denotes the number of clusters
in the system). Here $e^{\mu}$ is a fugacity which can
be used to tune the number of electrons in a grand-canonical
ensemble.

The expectation value of an operator $A$ is given by the usual
expression
\begin{equation}
\label{expecA}
\langle A \rangle = \frac{\sum_q A(q) \, W(q)}{\sum_q W(q)} \, .
\end{equation}
For the grand-canonical ensemble, the sum over
$q$ runs over all configurations of $n = 0, \ldots, \cal{N}$ electrons
while for the canonical ensemble, it is restricted to configurations
with a given number of electrons $n$. 

{}From the point of view of magnetism, a particularly important
observable is the square of the total spin ${\bm{S}}^2$
which can be written for a particular geometric configuration 
$q$ in two equivalent ways
\begin{equation}
\label{ssquared}
{\bm{S}}_{q}^2
= \sum_{i=1}^{M_q} \frac{\abs{C_i}}{2} \left(\frac{\abs{C_i}}{2} + 1\right)
= \sum_{l=1}^n {\cal{N}}n_q (l)\frac{l}{2}\left(\frac{l}{2}+1\right) \, .
\end{equation}
In the first form, the contribution from each cluster is manifest
while the second form relates it to $n_q(l)$, the normalized 
number of clusters of size $l$,
{\it i.e.}, a quantity which plays a central role in percolation theory
\cite{stauffer,isichenko}.

{\it{Quantum-statistical interaction.}}---Important differences arise between our 
Pauli-correlated percolation and the standard one [with trivial weight factor $W(q) \equiv 1$]. 
The weight factor can be cast as a pseudo-Boltzmann weight
of statistical origin, $W(q)\equiv\exp\left[\ln W(q)\right]$. The resulting effective entropic interaction, $\ln W(q)$, has the 
following properties. 
First, it is repulsive---a group of $m$ electrons has maximal weight $2^{m}$ if they 
form isolated one-electron ``clusters,'' and minimal weight $m+1$ if they form a single cluster. 
These extreme cases show that $\ln(m+1)\leq\ln W(q) \leq m\ln2$ 
saturates, {\it i.e.}, is never superextensive unlike other 
long-range interactions. Befitting its quantum statistical origin, the interaction is 
range-free---the shape of the cluster 
is unimportant, only its number of electrons matters.
Note also that the interaction is a genuinely many-body one: 
due to the form $\ln W(q)=\ln(m+1)$ it 
cannot be written as a sum of two-particle terms.

Taking all of this together demonstrates that this interaction gives rise to an
entirely novel ``Pauli-correlated'' percolation problem, of interest in 
its relevance to flat-band ferromagnetism and as a physically motivated 
example of a nonstandard percolation problem with an unusual weight. 

\begin{figure}
\centering
\includegraphics[width=\columnwidth]{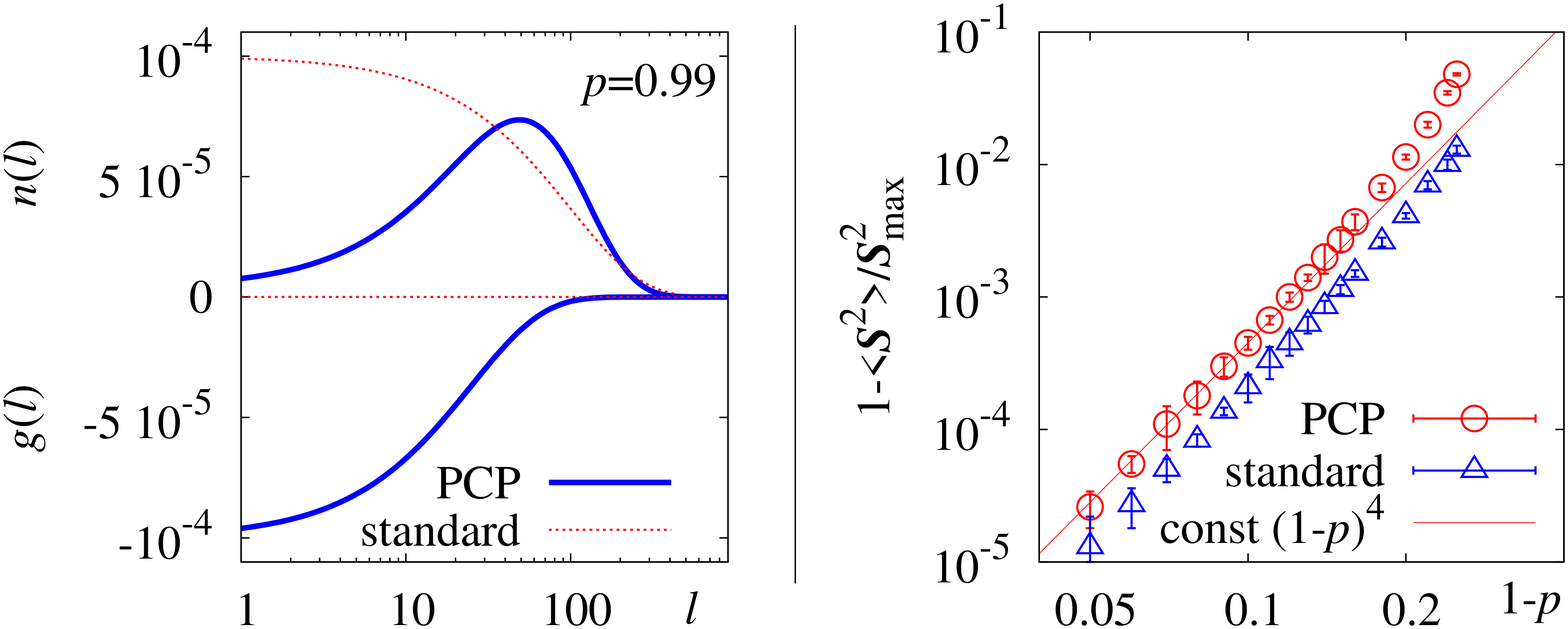}
\caption
{(Color online) Left: 
$n(l)$ (top) and $g(l)$ (bottom) at $p=0.99$ for PCP (solid line) and standard percolation (dotted line) in 1D. 
Right: Deviation of $\langle{\bm S}^2\rangle/{\bm S}^2_{\max}$ from saturation 
for large $p$ for PCP and standard percolation in 2D.}
\label{fig01a}
\end{figure}

{\it{Exact solution in one dimension.}}---We first 
provide a complete solution of the 1D Tasaki model (sawtooth chain) 
\cite{tasakiPRL,tasakiCMP,sawtooth1,sawtooth2, 1dtasaki_cerh3b2}.
A solution of the problem can be obtained with 
the help of a transfer matrix \cite{sawtooth2} despite the long-range nature of the statistical interaction
(for technical details see \ref{sec:SupplPcp}).
For a given electron density $p=n/{\cal{N}}$ we find
\begin{equation} 
\label{nl}
n(l)=\frac{4\,(1-p)^3}{(2-p)^2}\,(l+1)\,\alpha^l \; , \ \ \alpha = \frac{p}{2-p}~\, .
\end{equation} 
This cluster-size distribution (Fig.\ \ref{fig01a}, left panel) 
has a maximum at $l^\star > 1$ for $p>0.8$ moving along $l^\star\approx-(1+1/\ln \alpha)$ for $p\rightarrow1$.
This is  unlike the standard percolation result $(1-p)^2\,p^l$ \cite{stauffer}, which 
drops monotonically with $l$ and thus has a maximum at $l^{\star}=1$.

The macroscopic magnetic moment vanishes for $p < 1$, with a continuous onset at percolation, $p_f = 1$, as $(p_f - p)^{-1}$:
\begin{equation}
\label{s2}
\langle {\bm{S}}^2\rangle =  \frac{3\,p\,(2-p)}{8\,(1-p)} \, {\cal N}.
\end{equation}

The connected pair correlation function
\begin{eqnarray}
 g(\vert i-j\vert)&=&\langle n_in_j\rangle - \langle n_i\rangle\langle n_j\rangle \nonumber\\ 
&=&-(1-p)^2 \,e^{-\vert i-j\vert/\xi}<0 
\label{eq:G}
\end{eqnarray}
yields a correlation length $\xi=-1/(2\ln\alpha)$ 
that diverges as $(p_f-p)^{-1}$ when $p\to p_f=1$.
By contrast, for standard percolation there are no nontrivial pair
correlations: $g(\vert i-j\vert) = p \,(1 - p) \,\delta_{i,j}$.

The negative sign in Eq.~(\ref{eq:G}) shows
that the interaction is repulsive---electron positions 
anticorrelate. The long-range and many-body nature of the interactions leads to a nontrivial cluster 
size distribution favoring an approximately uniform spacing of vacant cells.

{\it{The phase diagram in 2D.}}---The 
2D case is not amenable to exact solution. Here we examine the 2D PCP numerically. 
Due to the nontrivial weights (\ref{degeneracy}), simple random sampling used for the conventional 
percolation is insufficient.

Going beyond standard numerical schemes \cite{hoshen,ziff} we have implemented efficient 
importance sampling on ${\cal L} \times {\cal L}$ square lattices with periodic boundary
conditions as follows. In the grand-canonical ensemble, we simply choose
a site and if it is empty (occupied), propose to insert (remove) an electron.
In the canonical ensemble we generate a new configuration $q_2$ from the given one $q_1$ by random permutation of two sites
in order to ensure a fixed number of electrons.
The new configuration is accepted with the
Metropolis probability $\min\left[1,W(q_2)/W(q_1)\right]$.
In addition, we have employed exchange Monte Carlo steps \cite{HN96}
for the grand-canonical simulations.
Clusters are identified in two different ways: 1) using a modified Newman-Ziff algorithm \cite{ziff} which locally updates cluster labeling 
for fixed number of occupied sites and 2) using the Hoshen-Kopelman algorithm \cite{hoshen} which makes a global update. 

Our central results
are the following.

{\em The percolation transition} is of first order as already suggested by visual 
inspection of individual configurations (Fig.\ \ref{fig01}). 
\begin{figure}
\begin{center}
\includegraphics[clip=on,angle=270,width=\columnwidth]{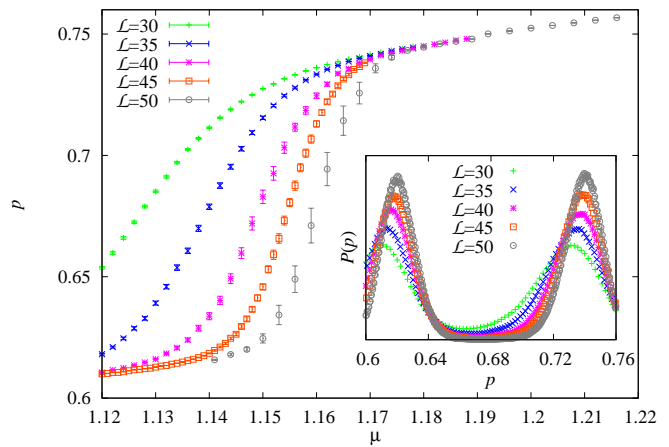}
\caption
{(Color online)
Density of electrons $p$ vs.\ 
chemical potential $\mu$, controlling the filling of the flat band in 2D.
Inset: Histograms of density for a ``finite-size'' critical value
$\mu=\mu_{c}$.}
\label{fig05}
\end{center}
\end{figure} 
Grand-canonical simulations exhibit a jump at a chemical potential $\mu_c$
between densities $p_-$ and $p_+$, 
fixed by equal-sized peaks in the histograms as shown in Fig.\ \ref{fig05}.
We estimate the jump to occur between densities 
$p_-$ around $0.63(1)$ and $p_+ \approx 0.75(2)$. In 
between, in our {\em canonical} 
simulations for finite systems, ferromagnetism appears to set on smoothly. Figure \ \ref{fig03} 
shows $\langle{\bm S}^2\rangle/{\bm S}^2_{\max}$ [where ${\bm S}^2_{\max}=\frac{n}{2}(\frac{n}{2}+1)$] for systems up to $270\times 270$ sites.
Additionally, the 
cluster-size distribution $n(l)$ indicates the emergence of a large
component
without passing through a scale-free critical distribution. 

\begin{figure}
\centering
\includegraphics[width=\columnwidth]{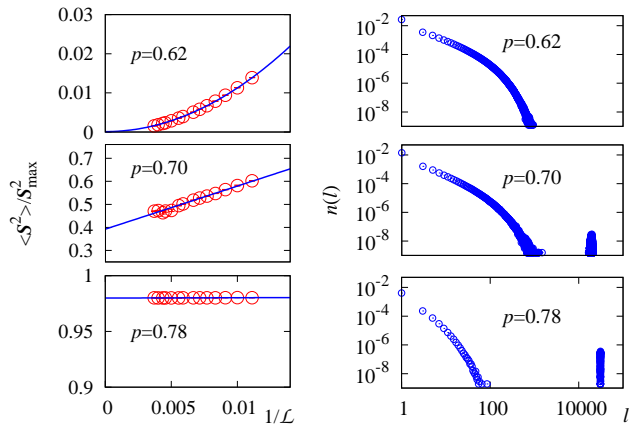}
\caption
{(Color online)
Left: Finite-size scaling of the square of the magnetic moment
$\langle{\bm S}^2\rangle/{\bm S}^2_{\max}$ for the 2D Tasaki lattice (up to ${\cal{L}}=270$).
Right: Normalized number of clusters $n(l)$ for ${\cal{L}}=200$.}
\label{fig03}
\end{figure}

{\em Extent of nonpercolating (paramagnetic) phase.}---The 
critical density for PCP exceeds that of the standard 
case ($p_{c} = 0.592746\ldots$ \cite{isichenko,ziff}), see Fig.\ \ref{fig03}, 
where the macroscopic moment at $p=0.62$ is seen to scale 
to zero with system size. This reflects the breakup of the large clusters due to the repulsive 
effective interactions.

{\em Percolating phase with unsaturated ferromagnetism.}---For 
higher densities, there appears a regime of {\em unsaturated} ferromagnetism, 
where $\langle \bm{S}^2\rangle/{\bm S}^2_{\max}
< 1$, illustrated by $p=0.78$ in Figs.\ \ref{fig01} and \ref{fig03}. The 
existence of this regime is transparent from the percolation viewpoint: In 
standard percolation the largest cluster excludes a nonzero density, 
$\mathcal{N}\,p\,(1-p)^c$ for $p \to 1$, $c$ being the coordination number of the lattice. 
For PCP, with its {\em repulsive} interactions, this is amplified 
(see right panel in Fig.\ \ref{fig01a}). However the power law is identical to that of standard
percolation, showing that the repulsive interactions do not 
immediately lead to a breakup of the largest cluster, presumably on account of the high entropic 
cost of arranging voids into continuous lines separating two clusters.

We note some features of the canonical ensemble arising in the phase-coexistence 
regime. The high-density phase appears to form first as a compact nonpercolating object with
a macroscopic magnetic moment [the configuration at $p=0.65$ shown in Fig.\ \ref{fig01}(d)
is in this region]. For higher densities, including the case $p=0.7$, the
ferromagnetic phase spans across the system [compare Fig.\ \ref{fig01}(e)].
The details of the phase-separated regime 
therefore contain much which is different from standard percolation
including the bootstrap and correlated variants \cite{CLR79,weinrib},
in particular with regard to properties which are of interest 
to flat-band ferromagnetism.
These topics are the subject of ongoing studies \cite{maksymenko}.

{\it{Conclusions and perspectives.}}---We 
have considered Pauli-correlated percolation, an unusual percolation problem
arising in a strongly correlated flat-band system, where the weights of
the geometrical configurations take nontrivial values due to
the spin degeneracy and the Pauli principle.

The Pauli-correlated percolation problem can be examined exactly 
in 1D and simulated efficiently in 2D.
We found that the effectively repulsive interaction leads to a
breaking up of the clusters, and thus to a first-order grand-canonical
transition in 2D, at a density which is higher than that of standard site
percolation.
For the underlying 2D Tasaki-Hubbard model our results imply ground-state ferromagnetism in a range of 
electron fillings from $0.21(1)$ to $1/3$.

Besides the 1D realization mentioned above \cite{1dtasaki_cerh3b2} and the hope of 
discovering corresponding quasi-2D materials, 
in \cite{organic_polymers} the possible realization of flat-band ferromagnetism in organic polymers 
was discussed. On the other hand 
the
2D version of the Tasaki lattice is so simple
that it seems to be a reasonable candidate
for realization as a system of cold atoms in optical lattices \cite{optical1,optical2}.

{\it{Acknowledgments.}}---The 
authors would like to thank J.~Chalker and A.~Nahum for valuable discussions. 
M.M.\ thanks DAAD (A/10/84322) and DFG (SFB 602)
for support of his stays at the University of G\"{o}ttingen 
and University of Magdeburg in 2010-2011.
A.H.\ acknowledges support by the DFG through a Heisenberg fellowship (Project HO~2325/4-2).
O.D.\ acknowledges the kind hospitality of University of Magdeburg in October-December of 2011.
J.R.\ and O.D.\ thank the DFG for support (RI 615/21-1).

\setcounter{secnumdepth}{3}

\setcounter{table}{0}
\setcounter{figure}{0}

\renewcommand{\thetable}{S\Roman{table}}
\renewcommand{\thefigure}{S\arabic{figure}}
\renewcommand{\thesection}{}
\renewcommand{\thesubsection}{S\arabic{subsection}}
 
\section*{Supplementary information}

\subsection{Pauli-correlated percolation and flat-band ferromagnetism}
\label{sec:SupplPcp}

These can be most transparently studied by considering the Hubbard model,
which describes electrons hopping on a lattice which
interact with each other via an on-site repulsion. The Hubbard Hamiltonian
reads
\begin{eqnarray}
\label{hamiltonian}
H=\sum_{\sigma=\uparrow,\downarrow}\sum_{\langle i,j\rangle}t_{i,j}
\left(c^{\dagger}_{i,\sigma}c_{j,\sigma}+{\rm{h.c.}}\right)
+U\sum_in_{i,\uparrow}n_{i,\downarrow}
\end{eqnarray}
in standard notations.

\begin{figure}[t!]
\begin{center}
\includegraphics[clip=on,width=7cm, angle=0]{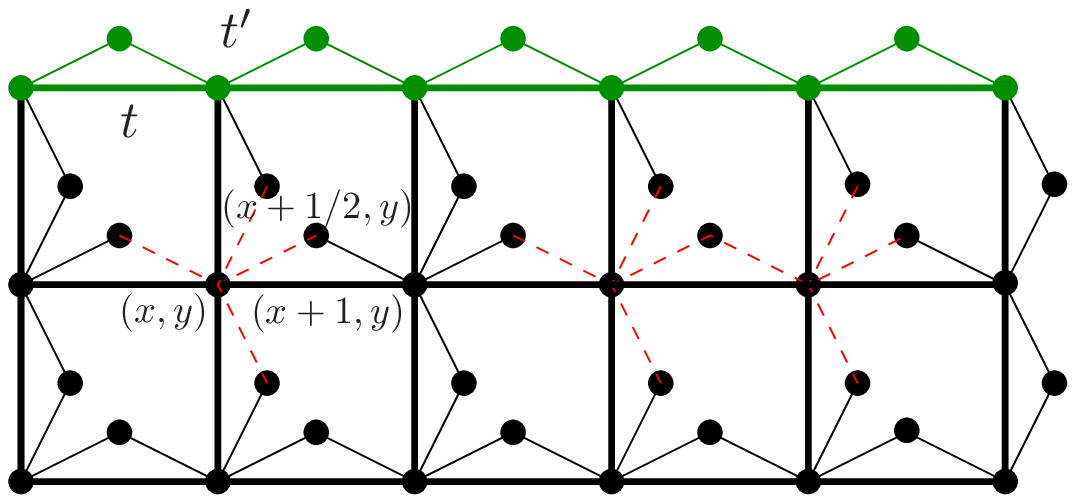}
\caption
{(Color online)
Two-dimensional Tasaki lattice [hopping integrals $t$ (thick lines) and
$t^{\prime}$  (thin
lines)].
A trapping cell contains five sites (dashed red lines).
The green circles  and lines show the 1D variant of the lattice (sawtooth chain).}
\label{fig01s}
\end{center}
\end{figure}

\begin{table*}[t!]
\begin{center}
\caption
{Ground-state degeneracies $g_{\cal{N}}(n)$ of the 2D Tasaki model,
as obtained by exact diagonalization for $U=\infty$ and $n$ electrons on $N=3\,{\cal N}$ sites. 
\label{tab1}}
\begin{tabular}[t]{|c||ccccc|cccc|cccc|} \hline
$\cal{N}$ & \multicolumn{5}{c|}{8}& \multicolumn{4}{c|}{10}& \multicolumn{4}{c|}{16} \\ \hline
$n$
& 1
& 2
& 3
& 4
& 5

& 1
& 2
& 3
& 4

& 1
& 2
& 3
& 4 \\ \hline
$E_n/t$
& $-4$
& $-8$ 
&$-12$ 
&$-16$ 
&$-20$ 

& $-4$ 
& $-8$ 
&$-12$ 
&$-16$ 

& $-4$ 
& $-8$ 
&$-12$ 
&$-16$  \\ \hline
$g_{\cal{N}}(n)$
&   16
&   96
&  256
&  372
&  336

&   20
&  160
&  640
& 1380

&   32
&  448
& 3584
&18008 \\ \hline
\end{tabular}
\end{center}
\end{table*}
Let us consider the Tasaki lattice
\cite{tasakiPRL,tasakiCMP} (Fig.~\ref{fig01s}) as a representative for a 2D 
system with a dispersionless (flat) band. Tasaki's lattice decoration
\cite{tasakiPRL,tasakiCMP} can be performed
in arbitrary dimension (and also for other lattices).
This allows a direct comparison of 1D and 2D.
For the lowest-energy one-electron band to be 
completely dispersionless,
the two relevant hopping integrals obey the relation
$t^{\prime}=\sqrt{c}\,t>0$, where $c$ is the coordination number of the 
underlying lattice.
The one-electron states in the flat band can be taken as localized on trapping cells.
In 2D each trapping cell consists of one site of the underlying square lattice and four neighboring decorating sites (Fig.~\ref{fig01s}).
A localized eigenstate of energy $\varepsilon_1=-c\,t=-4\,t$ of an electron with spin $\sigma$ is given by  
$l_{{\bm{r}},\sigma}^\dagger\vert 0\rangle$, 
where ${\bm{r}}=x,y$ runs over the sites of the underlying square lattice, 
$l_{{\bm{r}},\sigma}^\dagger=c_{x-\frac{1}{2},y,\sigma}^{\dagger}+c_{x+\frac{1}{2},y,\sigma}^{\dagger}
+c_{x,y-\frac{1}{2},\sigma}^{\dagger}+c_{x,y+\frac{1}{2},\sigma}^{\dagger}-2\,c_{x,y,\sigma}^{\dagger}$,
and $\vert 0\rangle$ denotes the vacuum state.

Localized  many-particle ground states for $n>1$ electrons and $U>0$ require (i) that each cell be empty or
singly occupied and (ii) that electrons in adjacent cells be in a symmetric
collective spin state. As a result, the spin degeneracy for a contiguous
cluster of $m$ electrons is reduced from $2^m$ to that of a spin $S = {m}/{2}$ multiplet, $m + 1$.

This is all that's needed for the mapping to PCP: all ground states can be labelled by the possible geometric configurations of $n$ electrons 
distributed over $\cal{N}$ cells (they are labelled by $q = 1,\, \dots,\, {{\cal{N}}  \choose n}$), and the
non-trivial weight of each state is given by Eq.~(\ref{degeneracy}),
where $M_q$ denotes the number of separated clusters, and $\abs{C_i}$ denotes the number of electrons in cluster $i$.

We have verified by 
exact diagonalization that this set of
localized many-body states spans the ground-state space for Tasaki lattices up to $N=3\,{\cal N}=48$ sites.
Numerical results for degeneracies are given in Table \ref{tab1} and agree with the corresponding
number of geometric configurations. Note that even for
the small lattices considered in this context, not all sectors up to $n = {\cal N}$
electrons are numerically accessible. In addition, the Hubbard model allows for
higher electron fillings which are beyond the the geometric picture of the present
paper.

\subsection{Pauli-correlated percolation in one dimension}
\label{sec:Suppl1D}

One-dimensional PCP can be analyzed using a transfer-matrix method. 
The crucial step is to choose suitable 
representative configurations for the quantum states with different values of $S^z$, 
for example by putting all the down spins right from the up 
spins in a cluster \cite{sawtooth2}.

We set $z=\exp(\mu)$ and
introduce the grand partition function $\Xi(z,{\cal{N}})$ of a percolating system
as the sum of probabilities of all possible random realizations.
$\Xi(z,{\cal{N}})$ will be written in terms of the transfer matrix ${\bf{T}}$ as
$\Xi(z,{\cal{N}})={\rm{Tr}}{\bf{T}}^{\cal{N}}$.
The transfer matrix for the Pauli-correlated percolation in one dimension then 
reads \cite{sawtooth2}
\begin{eqnarray} 
\label{transfer}
{\bf{T}} 
=
\left(
\begin{array}{ccc}
T(0,0) & T(0,\uparrow) & T(0,\downarrow)\\
T(\uparrow,0) & T(\uparrow,\uparrow) & T(\uparrow,\downarrow)\\
T(\downarrow,0) & T(\downarrow,\uparrow) & T(\downarrow,\downarrow)
\end{array}
\right)
=
\left(
\begin{array}{ccc}
1 & 1 & 1\\
z & z & z\\
z & 0 & z
\end{array}
\right).
\end{eqnarray}
The matrix elements $T(n_i,n_{i+1})$ correspond to the pair of neighboring sites $i$ and $i+1$
and acquire the value 1 ($z$) if the site $i$ is empty (occupied).

In order to determine the results in terms of $p$ rather than $z$,
we first calculate the average occupation number of the site $\langle n_i\rangle$ 
which should be equal to $p$.
Thus we have
\begin{eqnarray}
\label{av_oc-nu}
\langle n_i\rangle
=\frac{{\rm{Tr}{\bf{T}}^{\cal{N}}{\bf{N}}}}{\rm{Tr}{\bf{T}}^{\cal{N}}}
=p,
\;
{\bf{N}}=
\left(
\begin{array}{ccc}
0 & 0 & 0\\
0 & 1 & 0\\
0 & 0 & 1
\end{array}
\right).
\end{eqnarray}
In what follows we consider the thermodynamic limit ${\cal{N}}\to\infty$.
Using the left and right eigenvectors of ${\bf{T}}$, one obtains
$p(z) = {{1+4\,z-\sqrt {1+4\,z}}/[{1+4\,z}}]$, and inversion of this
function leads to $z(p)=p\,(2-p)/[4\,(1-p)^2]$.

We turn to the calculation 
of the average number of clusters of size $l$ (normalized by the lattice size ${\cal{N}}$) $n(l)$ \cite{stauffer}.
To fix the cluster of length $l$ 
we start with an empty site, 
then we have a string (cluster) of $l$ occupied sites,
and the last site of this string is followed by an empty one.
To calculate $n(l)$  
we have to replace the product of a sequence  of $l+1$ ${\bf{T}}$-matrices 
by the product ${\bf{S}}{\bf{C}}^{l-1}{\bf{F}}$, 
where
\begin{equation} 
\label{SCF}
{\bf{S}}=
\left(
\begin{array}{ccc}
0 & 1 & 1\\
0 & 0 & 0\\
0 & 0 & 0
\end{array}
\right),
\;
{\bf{C}}=
\left(
\begin{array}{ccc}
0 & 0 & 0\\
0 & z & z\\
0 & 0 & z
\end{array}
\right),
\;
{\bf{F}}=
\left(
\begin{array}{ccc}
0 & 0 & 0\\
z & 0 & 0\\
z & 0 & 0
\end{array}
\right).
\end{equation}
That yields
\begin{equation} 
\label{n_l}
n(l)
=\frac{{{\rm{Tr}}{\bf{T}}^{{\cal{N}}-l-1}{\bf{S}} {\bf{C}}^{l-1} {\bf{F}}}}{{\rm{Tr}}{\bf{T}}^{\cal{N}}}.
\end{equation} 
After straightforward calculations we arrive at
\begin{equation} 
\label{nla}
n(l)=\frac{4(1-p)^3}{(2-p)^2}(l+1)\alpha^l \; , \; \alpha = \frac{p}{2-p},
\end{equation} 
see Eq.~(\ref{nl}).
In these calculations we have used the relation 
\begin{equation} 
{\bf{C}}^m=
z^m\left(
\begin{array}{ccc}
0 & 0 & 0\\
0 & 1 & m\\
0 & 0 & 1
\end{array}
\right).
\end{equation}

Next, 
we use the transfer-matrix approach to calculate the (pair) site-occupation correlation function 
$g(l)=\langle n_in_{i+l}\rangle - \langle n_i\rangle\langle n_{i+l}\rangle=\langle n_in_{i+l}\rangle -p^2$. 
Using the matrix ${\bf{N}}$ defined in Eq.~(\ref{av_oc-nu}) 
and calculating ${\rm{Tr}}{\bf{T}}^{{\cal{N}}-l}{\bf{N}} {\bf{T}}^{l} {\bf{N}}$ 
we get 
\begin{equation} 
g(l)=-(1-p)^2 \alpha^{2\vert l\vert}.
\end{equation} 

Finally, 
we calculate the pair connectivity $\Gamma(n,n+l)$
(the probability that two sites $n$ and $n+l$ are both occupied and belong to the same cluster).
For this purpose we have to consider the quantity
\begin{eqnarray}
\label{connect}
\Gamma(n,n+l)
&=&\frac{{\rm{Tr}}({\bf{T}}^{{\cal{N}}-l}{\bf{N}}{\bf{C}}^l)}{{\rm{Tr}}{\bf{T}}^{\cal{N}}}\nonumber\\
&=&p\cdot\left(1+\frac{1-p}{2-p}l\right)\alpha^l.
\end{eqnarray} 

For completeness we mention that the transfer-matrix  approach can also be
applied to the standard percolation in one dimension. In this case we have
\begin{eqnarray} {\bf{T}}= \left( \begin{array}{cc} 1 & 1 \\ z & z \end{array}
\right), \; {\bf{N}}= \left( \begin{array}{cc} 0 & 0 \\ 0 & 1 \end{array}
\right), \nonumber\\ {\bf{S}}= \left( \begin{array}{cc} 0 & 1 \\ 0 & 0
\end{array} \right), \; {\bf{C}}= \left( \begin{array}{cc} 0 & 0 \\ 0 & z
\end{array} \right), \; {\bf{F}}= \left( \begin{array}{cc} 0 & 0 \\ z & 0
\end{array} \right) \end{eqnarray} and formulas (\ref{av_oc-nu}), (\ref{n_l}),
and (\ref{connect}) yield $z=p/(1-p)$, $n(l)=(1-p)^2\,p^l$,
$g(l)=p\,(1-p)\,\delta_{l,0}$, and $\Gamma(n,n+l)=p\cdot p^l$, respectively. These
results can of course also be obtained by more elementary means and are well-known in
standard percolation theory \cite{stauffer}.

In Fig.~\ref{fig01a} we illustrate $n(l)$ and $g(l)$ for both types of percolation at $p=0.99$.

\end{document}